*Article*

# Hyperspectral Imaging to detect Age, Defects and Individual Nutrient Deficiency in Grapevine Leaves

**Sourabhi Debnath** [1], **Tanmoy Debnath** [1], **Manoranjan Paul** [1,3,*], **Suzy Rogiers** [2], **Tintu Baby** [3], **DM Motiur Rahaman** [3], **Lihong Zheng** [1,3] and **Leigh Schmidtke** [3]

[1] Computer Vision Laboratory, School of Computing and Mathematics, Charles Sturt University, Australia;
Emails: sdebnath@csu.edu.au; tdebnath@csu.edu.au; mpaul@csu.edu.au; lzheng@csu.edu.au;
[2] NSW Department of Primary Industries, Australia; Email: suzy.rogiers@dpi.nsw.gov.au;
[3] National Wine and Grape Industry Centre, Charles Sturt University, Australia;
Emails: drahaman@csu.edu.au; tbaby@csu.edu.au; lschmidtke@csu.edu.au;
* Correspondence: mapul@csu.edu.au; Tel.: +61 2 63384260.

**Abstract:** Hyperspectral (HS) imaging was successfully employed in the 380 nm to 1000 nm wavelength range to investigate the efficacy of detecting age, healthiness and individual nutrient deficiency of grapevine leaves collected from vineyards located in central west NSW, Australia. For age detection, the appearance of many healthy grapevine leaves has been examined. Then visually defective leaves were compared with healthy leaves. Control leaves and individual nutrient-deficient leaves (e.g. N, K and Mg) were also analysed. Several features were employed at various stages in the Ultraviolet (UV), Visible (VIS) and Near Infrared (NIR) regions to evaluate the experimental data: mean brightness, mean 1st derivative brightness, variation index, mean spectral ratio, normalised difference vegetation index (NDVI) and standard deviation (SD). Experiment results demonstrate that these features could be utilised with a high degree of effectiveness to compare age, identify unhealthy samples and not only to distinguish from control and nutrient deficiency but also to identify individual nutrient defects. Therefore, our work corroborated that HS imaging has excellent potential as a non-destructive as well as a non-contact method to detect age, healthiness and individual nutrient deficiencies of grapevine leaves.

## 1. Introduction

Over the ages, researchers in Computer Science and other disciplines have utilised various image-based (including but not limited to hyperspectral (HS) image methods) experimental and analytical techniques for performing investigations with different fruits and vegetables. Diseases of pomegranate, betel vine and fungal infections in plants were detected using image processing [1-3]. Spectral reflectance methods have been applied for early disease detection in wheat fields and to assess the effect of leaf age and psyllid damage of *Eucalyptus saligna* foliage [4, 5]. Additionally, head blight contamination in wheat kernels was detected by using multivariate imaging of Fusarium [6]. Detection of multi-tomato leaf diseases (late blight, target and bacterial spots) in different stages using a spectral-based sensor are also reported [7]. Early and late blight diseases in tomato leaves, Fusarium head blight in wheat kernels, early detection of tomato spotted wilt virus and stress were detected using HS imaging [6, 8-11].

HS imaging is often related to high spectral and spatial resolutions and is a widely used technique for studies in disease and defect detection of leaves. It collects both spatial and spectral information simultaneously from an object in a non-destructive and non-contact way between UV and IR regions. As a result, its high spectral resolution provides an extensive volume of information in recognising, classifying and measuring objects [12, 13]. Change in spectral reflectance of wheat and lettuce leaves have been studied in response to micronutrient deficiencies using HS data [14, 15]. In these literature wheat plants with a different nutrient deficiency were compared with that of growing in the controlled environment with the aid of HS data. But the comparison between various nutrient deficiencies was not reported.

Grapes are one of the most widely grown berries in the world due to their nutrition value and importance in the multibillion-dollar wine industry, thereby related to thousands of jobs worldwide, and hence is an active area of research [16]. Grapevine leaves are a source of biomolecules which influence the quality and quantity of grapes. Photosynthetic performance of grapevine leaves is a necessary process, both qualitatively and quantitatively, for producing fruits [17]. The Photosynthesis depends on the intensity of light, temperature as well as chlorophyll, carotenoids and other accessory molecules such as nitrogen, and protein [18-20]. The contents of chlorophyll, as well as other molecules, change during the life cycle of leaves. In the early stage of development contents of chlorophyll, nitrogen and protein are high in leaves. Still, after a period when leaves enter a senescence phase, they gradually lose chlorophyll and protein. On the other hand, the health of leaves also affects photosynthetic processes as cellular structures of unhealthy leaves change, thereby influencing grape production [21-23]. Therefore, determining leaf age and detecting various nutritional defects in grapevine leaves will provide insight into vine performance and decision support tools for improved vineyard management. Few research reports are currently available that demonstrate the analysis of grapevine leaves utilising HS imaging. This work presents an extensive experimental study to detect age, healthiness and individual nutrient deficiencies of grapevine leaves with the aid of HS imaging. The primary contributions of this manuscript are:

- to predict age of grapevine leaves through experimental study of leaf surfaces,
- to determine effective wavelength region for age detection,
- to distinguish unhealthy from healthy leaves in terms of both visually defective leaves and leaves with nutrient deficiencies,
- to determine appropriate wavelength region for detection of visually imperfect leaves,
- to introduce a new metric called the variation index that indicates the deviation of a leaf's brightness compared to a matured healthy leaf labelled as the benchmark leaf as well as the control leaf which was employed for both age and unhealthy leaves detection studies,
- to distinguish between control leaf and nutrient deficient leaves, and
- to identify nutrient-deficient leaves individually (i.e. single nutrient deficiency).

This article is an expansion of our conference paper entitled *Detection of Age and Defect of Grapevine Leaves Using Hyper Spectral Imaging* [24] published in the Pacific-Rim Symposium on Image and Video Technology (PSIVT 2019) conference. The organisation of this paper follows this structure: section 2 details the methods, experimental setup and procedures, section 3 presents the experimental results, section 4 elaborates the details discussion of the experimental results and section 4 describes the conclusion and proposed future direction of this work.

## 2. Methods and Samples

*2.1.Experimental Setup*

A number of healthy and unhealthy grapevine leaves were collected from a vineyard located in central west New South Wales (NSW), Australia. A matured and healthy leaf, labelled as the benchmark leaf was selected as a reference leaf by visual inspection. Those leaves with discoloration and necrotic areas were catergorised as unhealthy. They are presented in Figure 1(b). For the nutrient deficiency study, grapevine leaves were collected from the National Wine and Grape Industry Centre (NWGIC) research centre, Charles Sturt University (CSU), NSW, Australia. They were scanned to generate HS data cubes on the same day of procurement by utilising Resonon's benchtop Hyperspectral imaging system available in the Computer Vision research laboratory at CSU. The HS imaging system is shown in Figure 1(a). It is comprised of a Pika XC2 benchtop high-precision HS camera, linear translation stage, mounting tower, lighting assembly, and software control system known as SpectrononPro. The imaging spectrometers are line-scan imagers, i.e. they collect data one line at a time. The Pika XC2 HS camera is Resonon's highest precision UV, VIS and NIR imager covering the ~380 - 1000 nm spectral range. This wavelength range could be divided into three regions: UV: ~380 nm to 400 nm, VIS: ~400+ nm to 700 nm and NIR ~700+ nm to 1000 nm.

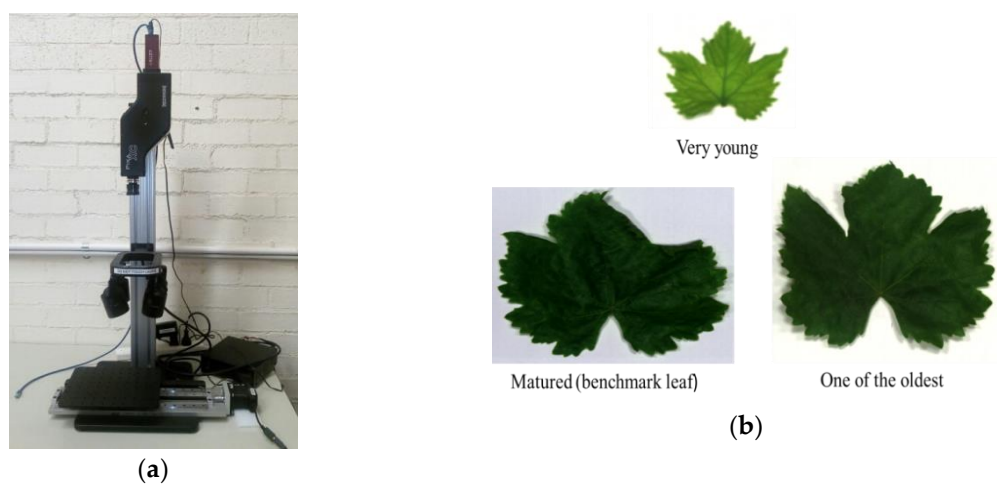

(**a**)

(**b**)

**Figure 1.** (**a**) Hyperspectral camera; (**b**) Samples of grapevine leaves according to age, including the benchmark leaf.

Multiple lines are scanned to assemble a complete two-dimensional image as the object is translated. The multiple line-images are then assembled line by line to form an entire image. Two-dimensional images are constructed by translating the sample relative to the camera. Sample leaves are placed on a linear translation stage. The stage needs to be in its leftmost position for an efficient scanning experience. The dark current has been acquired in the absence of a light source. After collecting multiple dark frames, SpectrononPro then uses these measurements to subtract the dark current noise from the analyses. Measuring absolute reflectance of an object requires calibration to account for illumination effects. Hence, a white reflectance reference was placed on the stage under light for a scan of the reference material. The collected data is then scaled in reflectance to the reference material, including flat-fielding to compensate for any spatial variations in lighting. Once the imager is calibrated for both dark current and reflectance reference, the imager will remain calibrated until the references are removed by the user. Within the camera settings, the frame rate and integration time were selected as 62.04244 Hz (default) and 244.01 ms, respectively. The speed unit and scanning speed were set to be linear and 0.07938 cm/s (lowest).

All samples were scanned for 10000 lines. The settings for the above-mentioned parameters were selected by trial and error to obtain the best resolution and HS image quality. Each leaf scan required approximately 5 minutes to generate an HS image file with a .bil extension and an approximate size of 1 GB. They were processed by SpectrononPro software. Spectra for each leaf were taken for ~380 nm to 1000 nm. At first, both sides of the leaves were scanned, but analysis suggested that there was not much difference in characteristics between the upper and lower surfaces of the leaves. Hence, the results reported here are based on the top (adaxial) surface of the leaves.

*2.2 Features*

For age detection studies, the experiments were conducted on a number of healthy grapevine leaves according to ascending order of their age, i.e. the youngest to the oldest leaf. Leaf age was based on the position on the shoot with the oldest leaves at the base and the youngest leaves at the growing point. To detect the age of grapevine leaves the following features were employed: mean brightness, mean 1st derivative brightness and variation index, which is a novel parameter introduced in this report. For a further study of detection of age, a proposed novel parameter, variation index ($v_i$) has been defined from the SD of brightness as:

$$v_i = \frac{(\sigma_{\text{benchmark leaf}} - \sigma_i) \times 100}{\sigma_{\text{benchmark leaf}}}\%$$ (1)

where $v_i$ is the variation index of the $i^{th}$ leaf, $\sigma_{\text{benchmark leaf}}$ and $\sigma_i$ are the SDs of the benchmark leaf and $i^{th}$ leaf respectively. This index describes how much the brightness of a leaf deviates from that of the benchmark leaf.

The differences between healthy and unhealthy leaves in all the three wavelength regions (UV, VIS and NIR) were probed. Both whole and selective areas have been studied to obtain an optimum selection method for distinguishing both types of leaves. The data cubes were obtained by selecting both the whole and the selective regions of leaves using the Lasso tool of SpectrononPro software to detect features. The characteristics features of leaves were studied in terms of mean 1st derivative brightness, mean spectral ratios and variation index. The mean brightness spectra for all the unhealthy leaves were each divided by the mean brightness spectrum of the representative healthy leaf to obtain mean spectral ratios, i.e. the benchmark leaf for both whole and selective areas of unhealthy leaves.

To compare the features of specific nutrient deficiencies a number of control leaves and leaves with Potassium (K), Magnesium (Mg) or Nitrogen (N) deficiency were studied using HS imaging. Here healthy leaves are represented as the control and unhealthy leaves by purposely induced nutrient deficiencies. Plants were grown in a temperature-controlled glasshouse for K and Mg deficiency. The N deficient leaves came from plants outdoors grown in a birdproof enclosure in NWGIC. The leaves were randomly collected from potted grapevines grown with specific nutrient regimes to induce K deficiency, Mg deficiency, N deficiency, and control plants with full nutrition. The nutrient regimes were developed by modifying Hoagland's nutrient solution. Table 1 presents a comparison of the leaves.

**Table 1.** Comparison of elemental concentration between the deficient group and the control plants. Elemental concentration is expressed as a percentage of dry matter in the leaf laminae.

| Element of interest | Deficient group | Control group | *P*-value |
|---|---|---|---|
| K% | K deficiency 0.3[a] | 4.2[b] | < 0.0001 |
| Mg% | Mg deficiency 0.1[a] | 0.6[b] | <0.0001 |
| N% | N deficiency 1.1[a] | 3.6[b] | <0.0001 |

In the table, values are averages of eight replicate leaves. Different letters show significant differences between the deficient group and the control group. Prism 8 (Graph Pad Software) was used for the statistical analysis. A significant difference was determined by t-test, *P* values <0.01 were considered to be substantial. The data cubes have been obtained by selecting the whole area of nutrient-deficient leaves using the Lasso tool of SpectrononPro software. The collected data of both control leaves and nutrient deficient leaves were then averaged. This deficiency study is reported in terms of mean brightness, normalised difference vegetation index (NDVI) [25-28], mean 1st derivative brightness, SD of brightness and variation index. This NDVI quantifies vegetation by measuring the difference between near-infrared (which vegetation strongly reflects) and red light (which vegetation absorbs). NDVI always ranges from -1 to +1 where positive values indicate vegetated areas and negative values signify non-vegetated surface features such as water, barren, clouds, and snow. A higher positive value of NDVI refers to healthy and dense vegetation [26]. The NDVI is associated with the reflectance of photosynthetic tissues that are calculated for individual scenes by the following equation [26] where RED and NIR stand for the spectral reflectance measurements acquired in the red (visible) and near-infrared regions, respectively. It is defined as

$$NDVI = \frac{NIR-\text{RED}}{\text{NIR+RED}} \qquad (2)$$

The 1st derivatives spectra have also been used to identify and analyse the location and properties of the redshift as it provides more sensitive measures of stress than broadband methods [15].

## 3. Experimental Results

Sections 3.1, 3.2 and 3.3 present experimental results related to the detection of age, healthy and unhealthy leaves and specific nutrient deficiencies respectively.

*3.1. Age Detection of Leaves*

3.1.1. Mean Brightness

Figure 1(b) displays samples of grapevine leaves, including the benchmark leaf used for the age detection studies. Figure 2 shows the mean brightness spectra of grapevine leaves of different ages in the wavelength range of ~380 to 1000 nm. Figure 3 presents the ratio of brightness intensity of all grapevine leaves with benchmark leaf at 554.3 nm.

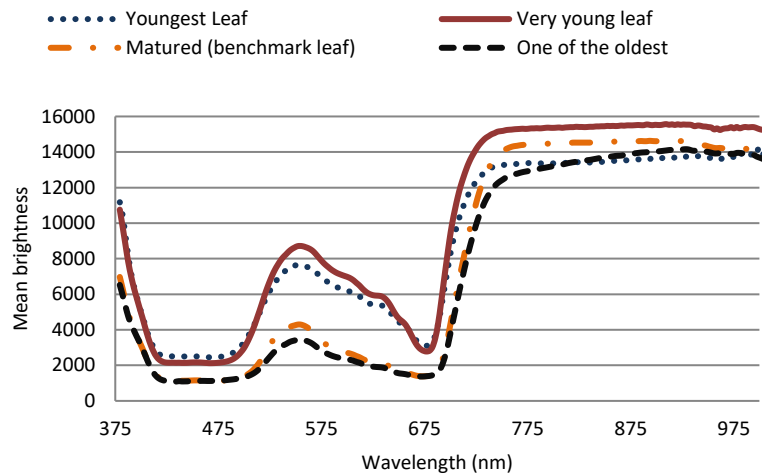

**Figure 2.** Mean brightness spectra of leaves of different age in the wavelength range of ~380 to 1000 nm.

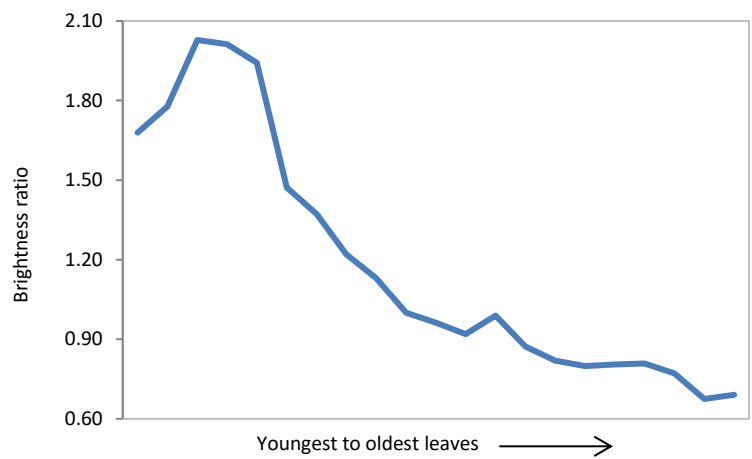

**Figure 3.** The ratio of brightness intensity of all grapevine leaves with benchmark leaf at 554.3 nm.

3.1.2. Mean 1st Derivative Brightness

Another parameter, namely mean 1st derivatives brightness providing information regarding the mean of 1st derivative brightness of pixels (i.e. within selected areas) in leaves, was employed to enhance our understanding of age detection. Figure 4(a) presents the mean 1st derivative brightness plot of leaves in the 400 nm to 675 nm range, and Figure 4 (b) demonstrates the maximum 1st derivative brightness peaks in the 400 to 675 nm range. Figure 5 represents the mean 1st derivatives brightness in the 675 nm to 775 nm range.

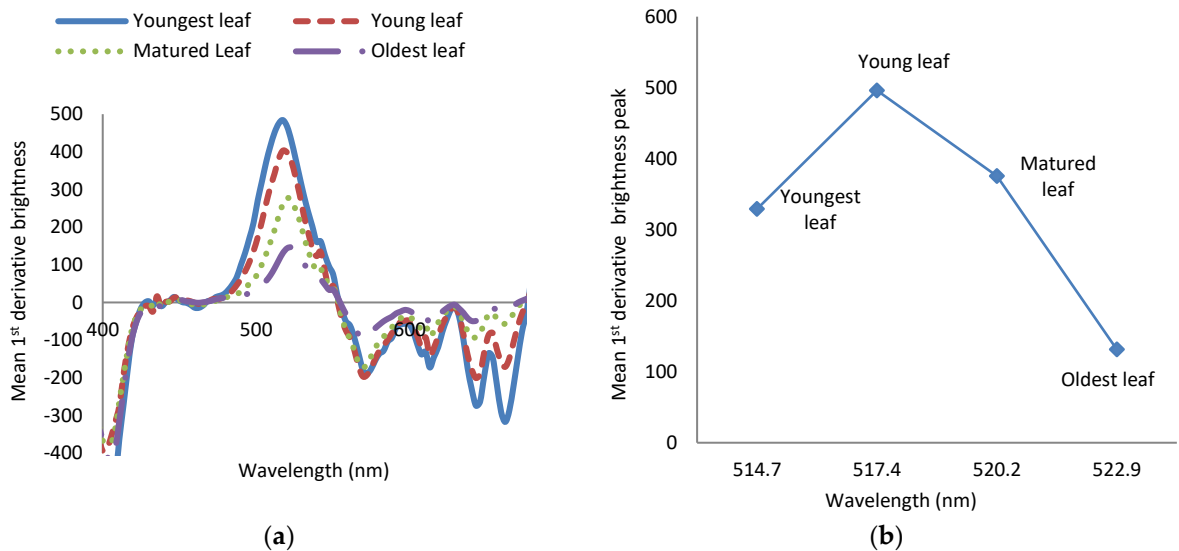

**Figure 4.** (**a**) Mean 1st derivative of brightness in the 400 nm to 675 nm range; (**b**) Maximum 1st derivative brightness peaks in the 400 to 675 nm range.

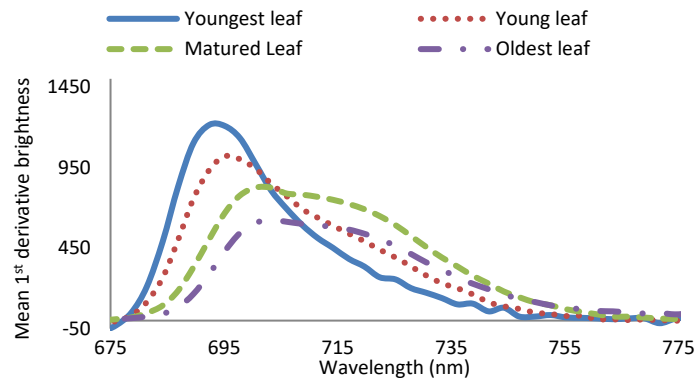

**Figure 5.** Mean 1st derivative of the brightness of leaves in the 675 to 775 nm range.

3.1.3. Variation Index

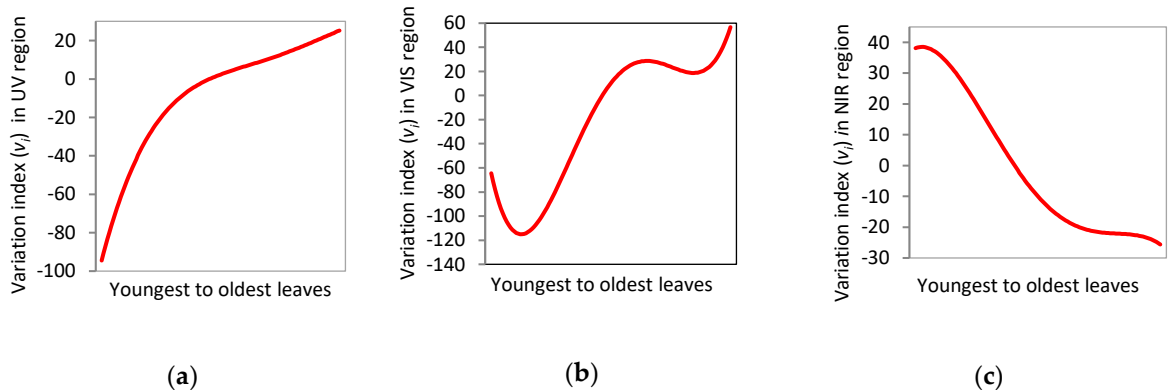

**Figure 6.** Variation index ($v_i$) of youngest to oldest leaves for determination of the age of leaves at: (**a**) UV region; (**b**) VIS region; (**c**) NIR region.

Figure 6 presents the plots of variation index (vi) vs age for the UV, VIS and NIR regions.

*3.2. Detection of Healthy and Unhealthy Leaves*

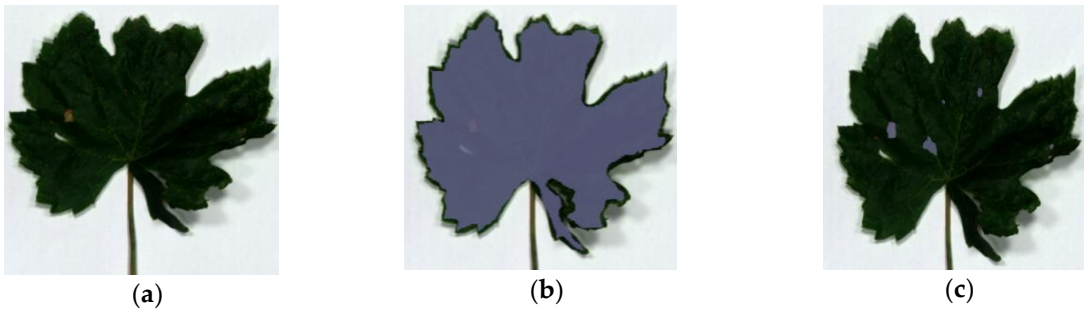

(a) (b) (c)

**Figure 7.** Unhealthy leaf with white spots and few brown spots and its selected areas for data cube accusation: (**a**) An unhealthy leaf with many small white spots and few small brown spots; (**b**) area (a grey area) selected for an unhealthy leaf; (**c**) Specific grey areas selected for an unhealthy leaf.

The following paragraphs detail the experimental results for the nine unhealthy leaves in terms of mean brightness and mean 1st derivative brightness *vs* wavelength. The characteristics of these leaves were also compared with a healthy leaf from the age detection group, i.e. the benchmark leaf (Figure 1(b)). This benchmark leaf was selected as it appeared to be a matured leaf. As seen in Figure 7(a), this particular unhealthy leaf contains white spots along with a few brown spots. Figures 7 (b and c) represent examples of the whole area and selective areas.

3.2.1. Mean 1st Derivative Brightness

Figure 8 presents the mean 1st derivative of the brightness of healthy (benchmark leaf), and unhealthy (whole and selective areas) leaves in the UV, VIS and NIR regions. Samples of unhealthy leaves with visible defects are presented in Figure 9.

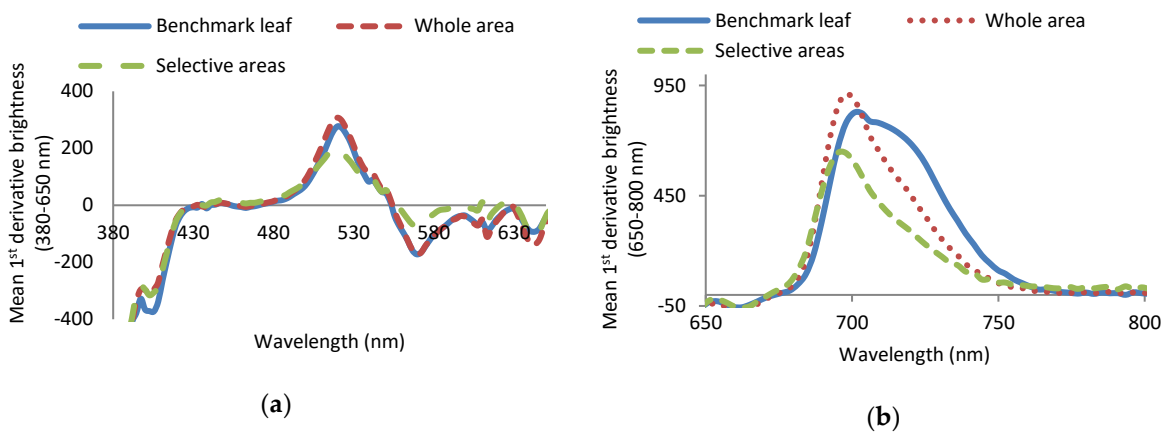

(a) (b)

**Figure 8.** Mean 1st derivative of the brightness of healthy (benchmark leaf) and unhealthy (whole and selective areas) leaves in the: (**a**) UV and VIS regions (380 nm to 650 nm); (**b**) NIR region (650 nm to 800 nm).

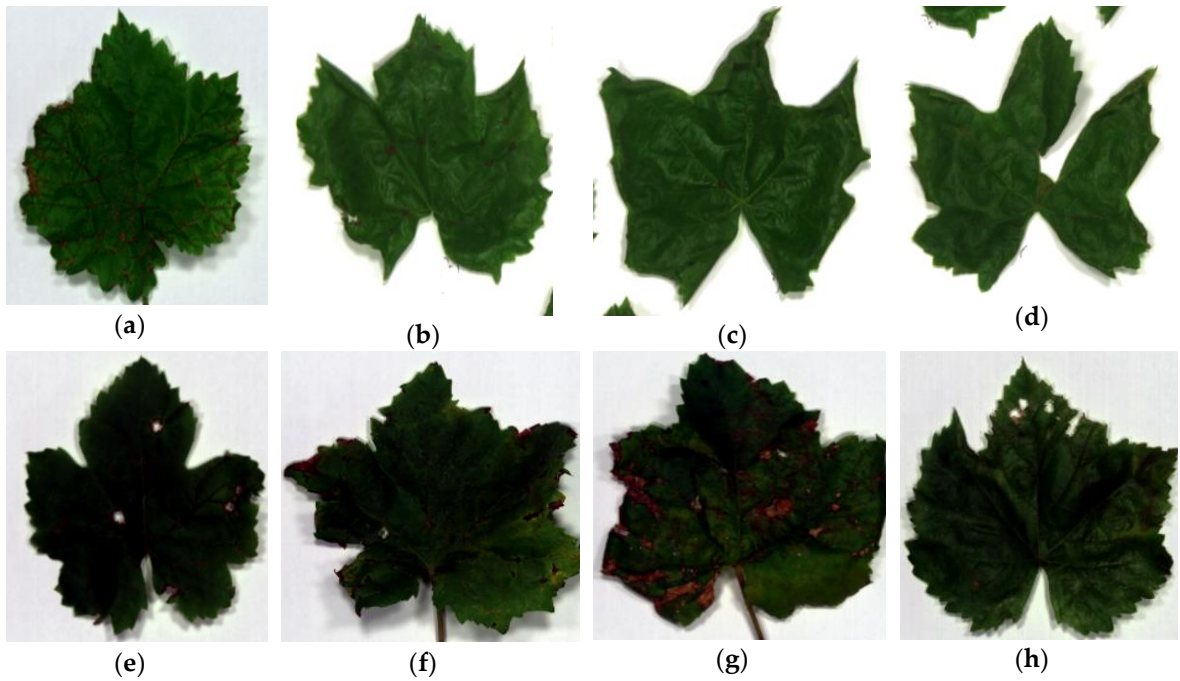

**Figure 9.** Sample of unhealthy leaves containing visible defects with: (**a**) brown areas and many small brown spots; (**b**) several big brown spots; (**c**) few brown spots; (**d**) some brownish areas; (**e**) few brown spots and some holes surrounded with brown regions; (**f**) brown and yellowish regions; (**g**) many large brown regions; (**h**) few brownish areas and holes surrounded by brownish areas.

3.2.2. Mean Spectral Ratios

Figure 10 presents mean spectral ratios of different unhealthy leaves (whole area) to benchmark leaf.

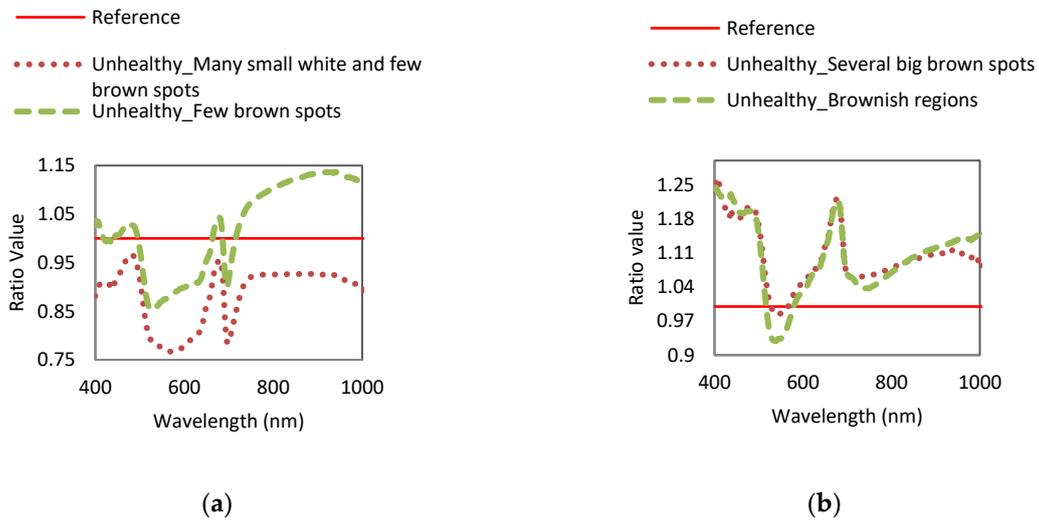

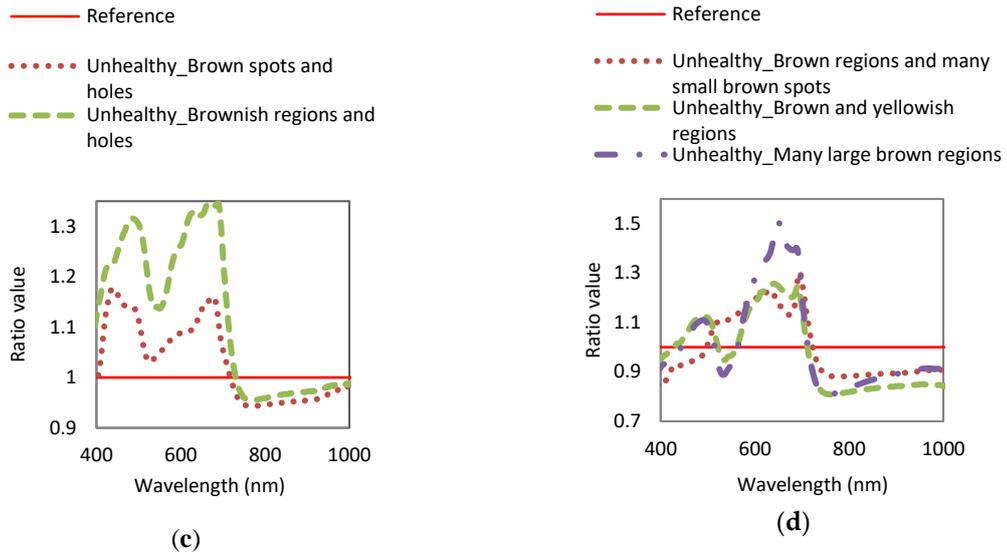

(c)   (d)

**Figure 10.** Mean spectral ratios of different unhealthy leaves (whole area) to benchmark leaf: (**a**) unhealthy leaves with many small white and few brown spots and unhealthy leaves with few brown spots; (**b**) unhealthy leaves with several big brown spots and unhealthy leaves with brownish regions; (**c**) unhealthy leaf with brown spots and holes and unhealthy leaf with brownish regions and holes; (**d**) unhealthy leaf with brown and yellowish regions and unhealthy leaf with many large brown regions.

### 3.2.3. Variation Index

Figure 11 demonstrated the Variation index ($v_i$) of healthy and unhealthy leaves in UV, VIS and NIR regions.

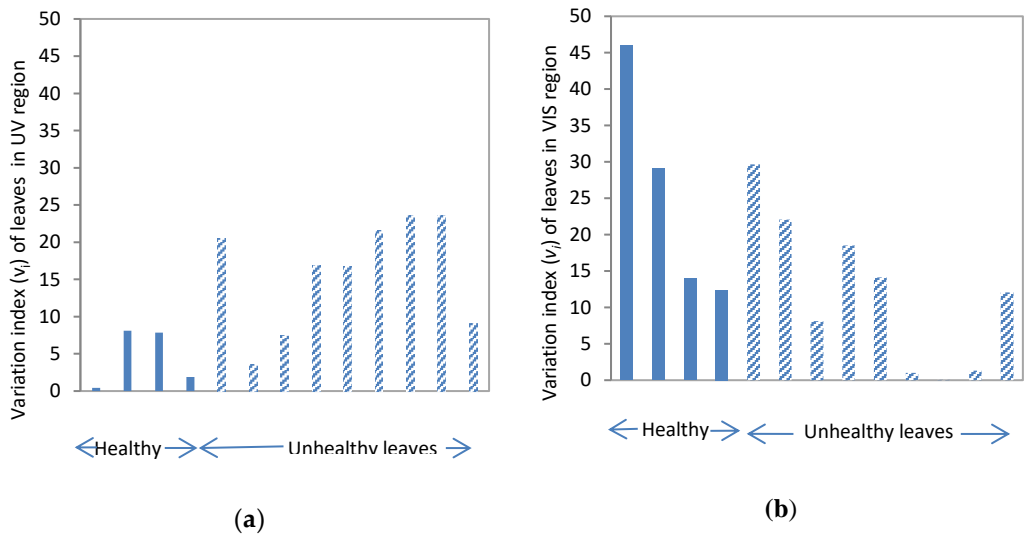

(a)   (b)

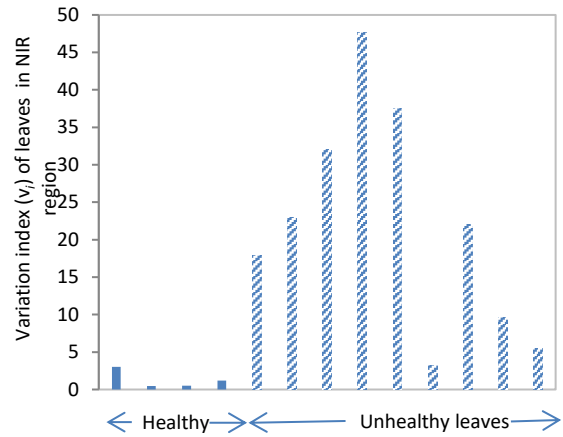

(**c**)

**Figure 11.** Variation index ($v_i$) of healthy and unhealthy leaves in the: (**a**) UV region; (**b**) VIS region; (**c**) NIR region.

*3.3. Nutrient Deficiency Detection*

3.3.1. Mean Brightness

Figure 12 shows the mean brightness spectra of the wavelength range from ~380 nm to 1000 nm of control and nutrient deficient leaves.

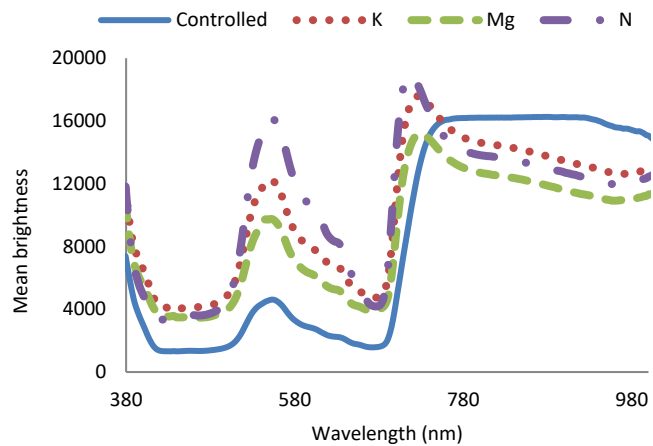

**Figure 12.** Mean brightness spectra of control and different nutrient-deficient leaves between ~380 nm to 1000 nm.

3.3.2. Mean Spectral Ratio

The ratio between the two entities was plotted in Figure 13 to compare the brightness of nutrient-deficient leaves with that of control leaves.

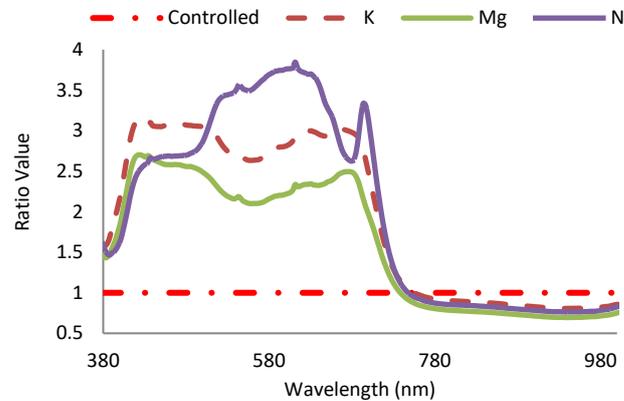

**Figure 13.** Mean spectral ratios of different nutrient-deficient leaves to control leaf for distinguishing control and different nutrient-deficient leaves.

3.3.3. Normalised Difference Vegetation Index (NDVI)

NDVI values of control leaves and nutrient deficient leaves are plotted in Figure 14.

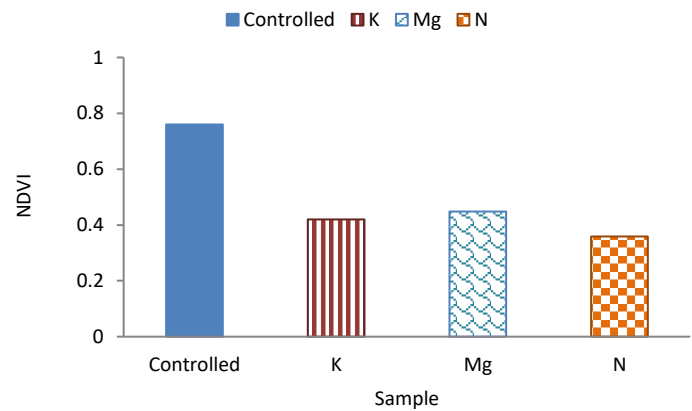

**Figure 14.** A plot of NDVI for control, K, Mg and N-deficient leaves.

3.3.4. Mean 1st Derivative Brightness

Figure 15 and 16 show the 1st derivatives spectra of control and nutrient deficient leaves within VIS wavelength and between 650 nm and 800 nm, respectively. From Figure 15, it is evident that in this wavelength range 1st derivative peak of nutrient-deficient leaves has shifted to the shorter wavelength (redshift) compared to the control leaves. It is known that this shift occurs if vegetation is exposed to low nutrients or environmental stress. Figure 17 presents a diagram of 1st derivative mean brightness ratio of control and specific nutrient-deficient leaves

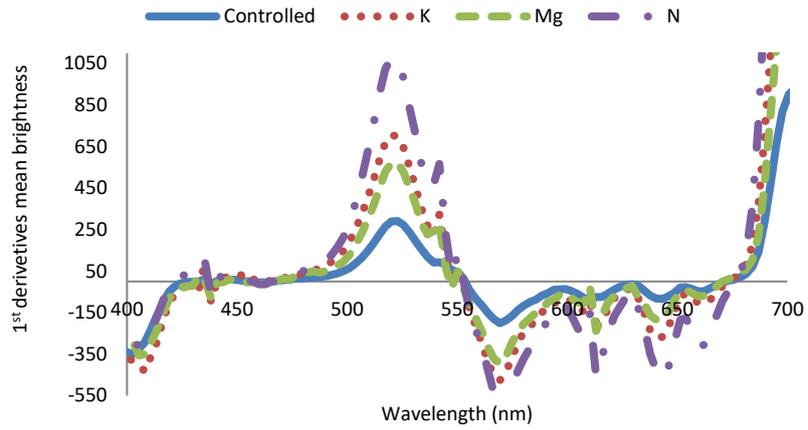

**Figure 15.** 1st derivative mean brightness spectra of control leaves and leaves with different nutrients deficiencies between VIS wavelengths.

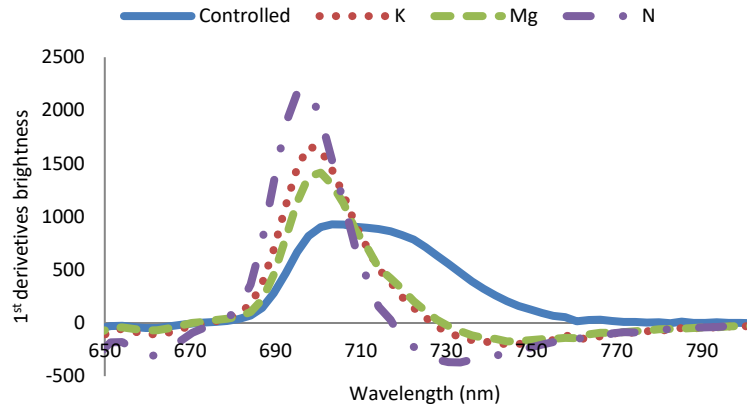

**Figure 16.** 1st derivative spectra of control and nutrient deficient leaves between 650 nm and 800 nm.

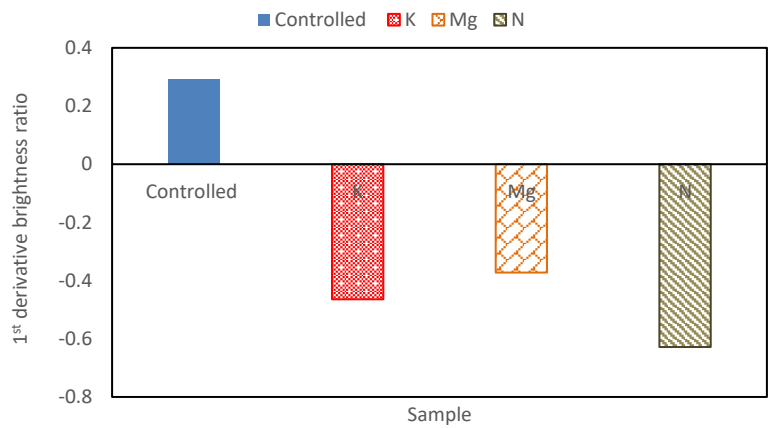

**Figure 17.** A plot of 1st derivative mean brightness ratio of control and different nutrient-deficient leaves.

3.3.5. Standard Deviation (SD)

Figures 18, 19 and 20 show the SDs of control and N, Mg or K deficient leaves in the UV, VIS and NIR regions of wavelength respectively.

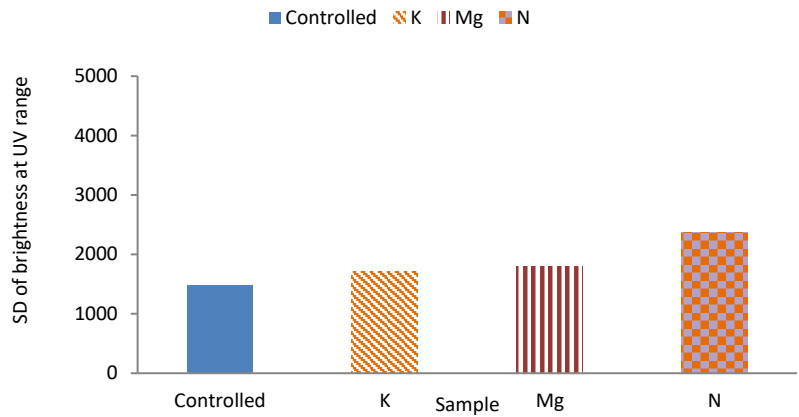

**Figure 18.** A plot of SD of control and different nutrient-deficient leaves in UV region.

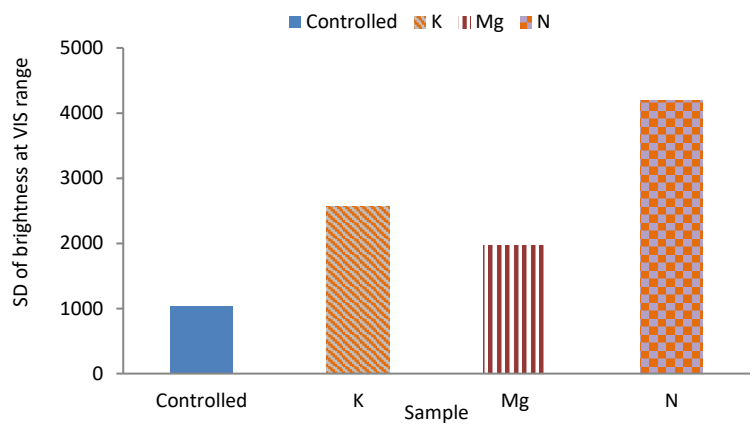

**Figure 19.** A plot of SD deviation of control and different nutrient-deficient leaves in the VIS region.

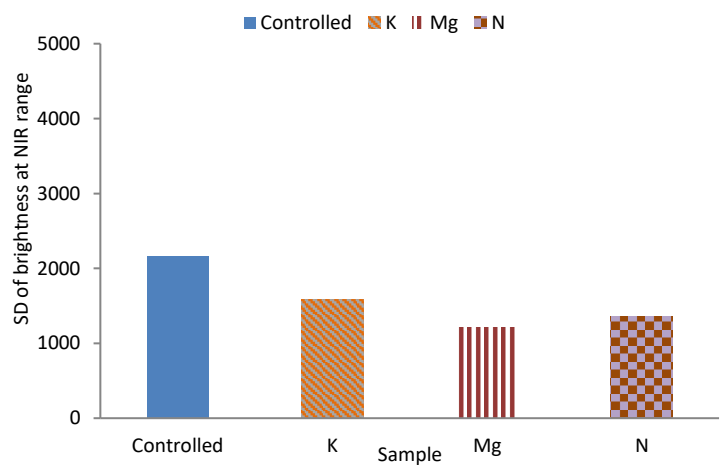

**Figure 20.** A plot of SD of control and different nutrient-deficient leaves in the NIR region.

3.3.6. Variation Index

Figures 21, 22 and 23 present the Variation indices ($v_i$) of leaves in the UV, VIS and NIR regions respectively for distinguishing control and different nutrient-deficient leaves.

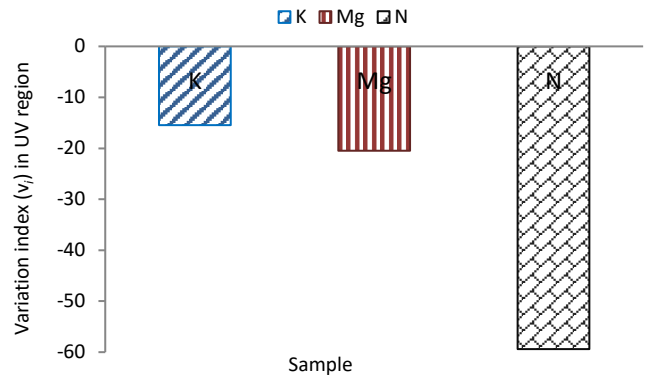

**Figure 21.** Variation index ($v_i$) of leaves in the UV region for distinguishing control and different nutrient-deficient leaves.

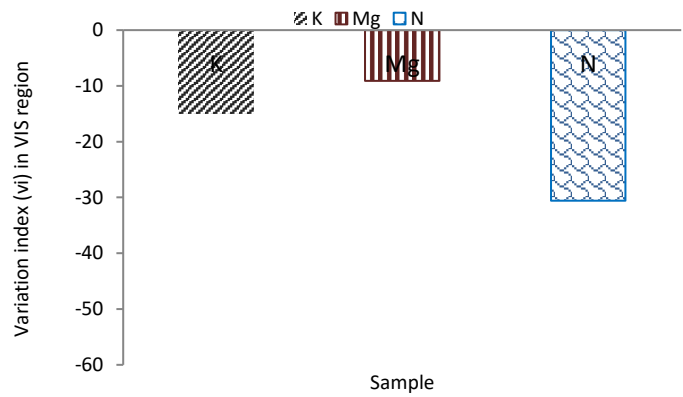

**Figure 22.** Variation index ($v_i$) of leaves in the VIS region for distinguishing control and different nutrient-deficient leaves.

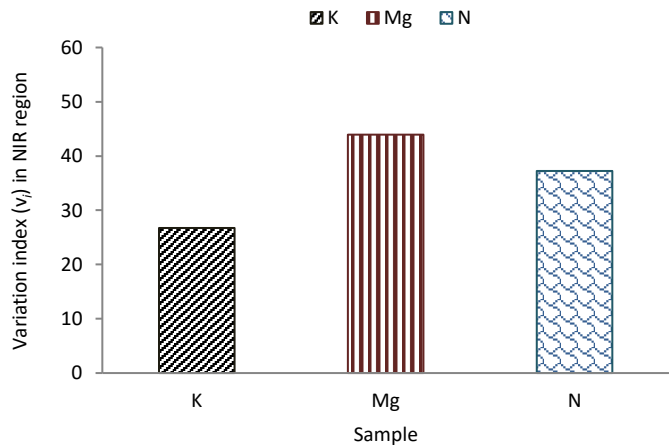

**Figure 23.** Variation index ($v_i$) of leaves in the NIR region for distinguishing control and different nutrient deficiencies.

## 4. Discussions

Sections 4.1, 4.2 and 4.3 elaborate discussions related to the experimental results presented for the detection of age, healthy and unhealthy leaves and nutrient deficiencies respectively in section 3.

*4.1. Age Detection of Leaves*

4.1.1. Mean Brightness

Figure 2 demonstrates that in the VIS range of spectra, a characteristic high brightness peak within the green wavelength range has been obtained. The characteristic curve is governed by the absorption effect from chlorophyll and other leaf pigments in the visible wavelength range. It absorbs blue and red wavelengths more strongly than green. As a result, it reflects a higher amount of green compared to blue and red and hence the characteristic high brightness [29-31]. It can also be observed that the magnitude of mean brightness of very young leaves is higher than those of youngest, benchmark and older leaves.

Figure 2 also corroborates that in the NIR region, the mean brightness of grapevine leaves is higher than that of the VIS region as a plant leaf has a low reflectance in the VIS spectral region due to strong absorption by chlorophylls and carotenoids. On the other hand, a relatively high reflectance in the near-infrared is observed because of internal leaf scattering and no absorption [30, 32]. Unfortunately, these mean brightness spectra from the UV to NIR ranges do not provide an unambiguous indication of leaf age. Hence the brightness of leaves was compared to that of benchmark leaf at 554.3 nm. Spectral analysis of experimental data reveals that the brightness difference between the oldest to the youngest leaves is maximum at ~ 554.3 nm.

From Figure 3, it has been found that the ratio of brightness of the first few young leaves increases. After that, it decreases with the increasing age of leaves at 554.3 nm. The phenomenon of increasing brightness ratio for the first few very young leaves could be due to the effect of water content. Though the chlorophyll content of younger leaves is lower than that of matured and older leaves [5], the water content is higher [33]. It is reported that a decrease in water content was found to increase reflectance [34]. Therefore, the level of the water content has a dominant effect on the reflectance of young leaves of a certain age. Hence reflectance of green wavelength, as well as brightness, is lowest for the youngest leaf. Figure 3 demonstrates that few very young leaves have a slightly lower brightness ratio than those of young slightly more mature leaves. It is also evident that the brightness ratio decreases with increasing maturity, but the decrement is much more gradual for the matured to oldest leaves. Expanding leaves (i.e. leaves that are growing in size, but not yet fully mature) combine high greenness with a low net positive photosynthetic rate due to some carbon inflow from other source areas within the vine. As leaves get older, beyond full-expansion, water content decreases, potentially due to lower stomatal conductance, thereby decreasing brightness. This phenomenon is caused by various effects such as leaf ageing [35,36]. The loss of chlorophyll also occurs in older leaves. Therefore, a decreasing brightness ratio could arise from a lower photosynthetic rate due to the ageing effect.

4.1.2. Mean 1st Derivative Brightness

Analysis suggests that the youngest, young, matured and oldest leaves have the highest mean 1st derivative brightness peaks at ~514 nm, ~517 nm, ~520 nm and ~523 nm respectively. Hence, a longer wavelength shift of the 1st derivatives mean brightness peaks were observed in the 470 nm to 550 nm range due to ageing as demonstrated in Figure 4(b). It is evident in Figure 5 that the brightness peaks started gradually broadening with age, i.e. with maturity, the mean rate of change of brightness becomes slower.

4.1.3. Variation Index

It is evident in Figure 6 that there are definite relational trends between variation index ($v_i$) and age in both UV and NIR regions. With increasing age of leaves, $v_i$ gradually increases and decreases in the UV and NIR regions respectively. However, in the VIS region, the relationship is not a straightforward one. Here $v_i$ first decreases, then increases and afterwards follows this pattern once again with age. Hence it might be stated that this parameter is an efficient predictor of age in the UV and NIR regions but not in the VIS region.

*4.2. Detection of Healthy and Unhealthy Leaves*

4.2.1. Mean 1st Derivative Brightness

Figure 8(a) states that other than changes in the magnitude of mean 1st derivative brightness, no significant information can be obtained for distinguishing benchmark, the whole area and selective area leaves in the UV and VIS wavelength regions.

Figure 8(b) demonstrates that the mean 1st derivative brightness in the NIR wavelength range is higher for unhealthy leaves compared to the benchmark leaf across the entire leaf area. Narrowing of the brightness curves for unhealthy leaves is evident for the whole area compared to benchmark leaf. A shorter wavelength shift of the peak is also observed for unhealthy leaves compared to benchmark leaf. Reflectance is insensitive to change in chlorophyll content but sensitive to internal leaf structure, water content, structural compounds and changes in internal mesophyll structure [38, 39] as in the case of the formation of necrotic areas due to tissue decay. Therefore, the deviation of the mean 1st derivative brightness curves of unhealthy leaves compared to benchmark leaf could be due to the difference in internal leaf structures.

The magnitude of mean 1st derivatives brightness for the selective areas is lower than the whole area case though characteristic curves are similar in appearance. For the selective areas case, areas have been selected based on visible spots or defective regions. Therefore, characteristic features of only visibly defective areas have been obtained. On the other hand, for the whole area case the whole area of a leaf has been selected regardless of visible defects. As a result, the features of both visible and nonvisible defective areas have been acquired. Therefore, the whole area case probes significant features compared to selective areas case.

For most of the unhealthy leaves, similar features to Figure 8, such as narrowing and shift of the curves towards shorter wavelengths were observed. It has been found that a sharp transition from low to high reflectance usually occurs in the wavelengths between the VIS and NIR regions, and this transition usually shifts to shorter wavelengths in diseased crops [39]. The wavelength where this transition occurs can be observed in the 1st derivative brightness curves, as presented in Figure 8(b). Hence, this deviation of 1st derivative brightness curve shapes could be due to the brown spots in unhealthy leaves. Analysis of the experimental data confirms that mean 1st derivative brightness parameter could significantly distinguish the healthy and most (though not all) unhealthy leaves in the NIR region but not so efficiently in the UV and VIS regions.

4.2.2. Mean Spectral Ratios

Unhealthy leaves with many small white and few brown spots and unhealthy leaves with few brown spots exhibit almost similar ratio curve trends from ~ 400 nm to 1000 nm as per Figure 10(a). Here the brightness ratio is higher for unhealthy leaves with few brown spots than that of unhealthy leaves with many small white and few brown spots.

From Figure 10(b) it can be surmised that unhealthy leaves with several big brown spots and unhealthy leaves with brownish regions have almost similar ratio curves. The brightness ratio of an unhealthy leaf with several big brown spots is more significant than an unhealthy leaf with brownish regions between the wavelength range ~ 491 nm and ~ 825 nm. After 825 nm, the brightness ratio is higher for the unhealthy leaf with brownish regions than the unhealthy leaf with several big brown spots.

Figure 10(c) presents that an unhealthy leaf with brown spots and holes and unhealthy leaf with brownish regions and holes have similarly shaped ratio curves. The brightness ratio of the former is higher than the later in the ~ 400 nm and 1000 nm wavelength range.

Figure 10(d) states that for whole area cases, the unhealthy leaf with brown and yellowish regions and unhealthy leaf with many large brown regions have similar ratio curves with higher ratio values for the later between ~ 570 nm to 1000 nm.   Between ~ 400 nm and ~ 570 nm, the ratio value is higher for unhealthy leaves with brown and yellowish regions than that of an unhealthy leaf with many large brown regions. On the other hand for the unhealthy leaf with brown regions and many small brown spots, the ratio curve is similarly shaped to unhealthy leaves with brown and yellowish regions and unhealthy leaf with many large brown regions between ~ 603 nm and 1000 nm. From ~ 400 nm to ~ 603 nm the unhealthy leaf with brown regions and many small brown spots ratio curve are different to unhealthy leaves with brown and yellowish regions and unhealthy leaf with many large brown regions as well as curves of other unhealthy leaves.

4.2.3. Variation Index

Variation index $v_i$ (equation 1) has also been applied to obtain a correlation between healthy and unhealthy leaves. In this case, $\sigma_i$ represents the SDs of the $i^{th}$ unhealthy leaf. Figure 11(c) demonstrates that healthy leaves have $v_i$ indexes closer to benchmark leaf, i.e. unhealthy leaves have higher value of $v_i$ indexes compared to healthy leaves at NIR region. No such prominent patterns were observed for unhealthy and healthy leaves in the UV and VIS regions, as presented in Figure 11(a) and Figure 11(b).

*4.3. Nutrient Deficiency Detection*

4.3.1. Mean Brightness

It can be observed from Figure 12 that in the VIS wavelength, the brightness of leaves with nutrient deficiencies is higher than the control leaves. Nutrient deficiency affects chlorophyll content and generally increases brightness in the VIS wavelength. The brightness peak is generally centred at ~ 550 nm and broadened towards the red as absorption of incident light by chlorophyll decreases [14]. It can also be seen from Figure 12 that N deficient leaves have higher brightness compared to control, K and Mg deficient leaves at ~550 nm. This effect may be because N is needed for chlorophyll production [40].

Figure 12 also states that the brightness peak of nutrient-deficient leaves has appeared at the shorter wavelength (redshift) compared to the control leaf in the wavelength range between ~ 680 nm and 1000 nm. The slope of the redshift is steeper for nutrient deficient leaves than that of the control leaves. When plants are under stress (e.g. nutrient deficiency), there is a loss of chlorophyll and hence chlorophyll absorption and NIR reflectance decrease due to structural changes in the plant. This phenomenon causes a shift in the red edge position to shorter wavelengths occurring between the areas of red and NIR [14, 15]. It can also be observed that N deficient leaves shifted more to the shorter wavelength compared to the Mg and K deficient leaves. It is evident that the brightness of nutrient deficient leaves are lower compared to the control leaves in the wavelength region between ~750 nm to 1000 nm. Here Mg has the lowest brightness compared to the other leaves.

4.3.2. Mean Spectral Ratio

Figure 13 demonstrates that the curve shapes for K and Mg deficient leaves are similar, but the magnitude of Mg deficient leaves is lower than that of K deficient leaves. The curve shape of N deficient leaves is quite different from those of K and Mg deficient leaves. There are peaks between ~650 nm and ~ 750 nm for N, Mg and K deficient leaves. In this wavelength region peaks at ~693 nm, ~680 nm and ~663 nm have been found for N, Mg and K deficient leaves respectively. It can also be found that magnitude of these peaks at wavelengths mentioned above is ~3.3, ~2.5 and ~3 times higher for N, Mg and K deficient leaves compared to the control leaves. All mineral deficiencies cause modifications in the brightness of leaves compared to the control specimens [14].

4.3.3. Normalised Difference Vegetation Index (NDVI)

From NDVI values of Figure 14, it can be observed that they are is significantly higher for control leaves than those of nutrient-deficient leaves. These values indicate that control leaves are healthier than leaves with N, K and Mg deficiencies. It can also be observed that the NDVI values for leaves with N, K and Mg deficiencies are not significantly different from each other.

4.3.4. Mean 1$^{st}$ Derivative Brightness

From Figure 15, it can be found that between ~460 nm and ~ 540 nm peak magnitudes of 1$^{st}$ derivatives of mean brightness is higher for nutrient deficient leaves compare to that of control leaves. In this wavelength region, the 1$^{st}$ derivatives brightness magnitudes are ~36%, ~19% and ~24% higher for N, Mg and K deficient leaves compared to control leaves respectively [15].

Figure 16 states that the ~700 nm derivative peaks are dominant in leaves with nutrient deficiencies, and these results are supported from published findings [41, 42]. Narrowing of 1$^{st}$ derivative brightness peak can also be observed for the nutrient-deficient leaves compared to control leaves. The midpoint of this broadened peak of control leaves is at ~ 708 nm. This broadening and narrowing of 1$^{st}$ derivative peaks of control and nutrient deficient leaves respectively could be due to the amount of chlorophyll content in the leaves. In this region, the 1$^{st}$ derivative peak magnitude is ~ 15%, ~ 9% and 7% higher for N, K and Mg deficient leaves respectively in comparison to control leaves.

The 1$^{st}$ derivative mean brightness shift can further be studied by using the HS derivative ratio in the boundary between the region of the strong absorption of red light by chlorophyll (~ 680 nm) and the region of high multiple scattering of radiation (~750 nm) [41]. From the 1$^{st}$ derivative ratio plot of Figure 17, it can be observed that this 1$^{st}$ derivative ratio is positive for control leaves and negative for nutrient deficient leaves. Furthermore, the ratio values are ~ 62%, ~47% and ~37% for N, K and Mg in the negative direction respectively. It is evident that similar to the UV region control leaves have the lowest SD value compared to different nutrient-deficient leaves. Here N deficient leaves have the highest value among these leaves. The difference between SD of K and Mg is 600 in the VIS region. This difference value is higher in the VIS region compared to UV.

In contrast to UV (Figure 18) and VIS (Figure 19) regions, the SD value is highest for control leaves than those of nutrient-deficient leaves in the NIR region (Figure 20). In this region, this value has also been found to be highest for N deficient leaves compared to K and M counterparts. Here the SD values of N, Mg and K are close to each other. The difference of SD values between N and Mg, N and K, and K and Mg were found to be 100, 300 and 400, respectively.

4.3.5. Standard Deviation (SD)

To distinguish features of control and nutrient deficient leaves, SD has also been studied for UV, VIS and NIR regions of wavelength. Figure 18 corroborates that the SD of control leaves in UV range has the lowest value compared to those of nutrient-deficient leaves. This value is highest for N deficient leaves than the control, K and Mg deficient leaves. But the difference of this SD value between K and Mg deficient leaves is ~ 100.

4.3.6. Variation Index

From Figure 21 it can be visualised that $v_i$ values are ~60%, ~21% and ~16% higher for leaves with N, Mg and K deficiency respectively in the negative direction in the UV region. This phenomenon indicates that the SD of leaves with a nutrient deficiency is higher than that of the control leaves. In the case of VIS region (Figure 22), it can be observed that $v_i$ values deviate ~31%, ~ 8% and ~15% for leaves with N, Mg and K deficiencies respectively in the negative direction indicating that $v_i$ of leaves with a nutrient deficiency is higher than that of the control leaves.

Furthermore, it can be obtained from Figure 23 that in NIR region $v_i$ values deviate ~38%, ~ 44% and ~27% for leaves with N, Mg and K deficiencies respectively in a positive direction indicating that here $v_i$ of leaves with nutrient deficiency is lower than that of control leaves. Table 2 below presents different essential features and corresponding values for enabling classification to distinguish healthy and nutrient deficient leaves.

**Table 2.** Summary of Features

| Feature | Control | N | Mg | K | Remark |
|---|---|---|---|---|---|
| NDVI | High (0.75) | Lowest (0.35) | Lower (0.45) | Lower (0.42) | Can be an excellent feature to distinguish control from nutrient deficient leaves. But individual nutrient deficiencies may not be uniquely identified. |
| Variation index ($v_i$) in UV region | - | High (59) | Lower (21) | Lowest (15) | Can be used as a feature to distinguish N deficiency from K and Mg deficiencies. But K and Mg deficiencies may not be uniquely identified. |
| Variation index ($v_i$) in the VIS region | - | High (31) | Low (9) | Moderate (14.5) | Can be used as a feature to distinguish N deficiency from K and Mg deficiencies. But K and Mg deficiencies may not be uniquely identified. |
| Variation index ($v_i$) in the NIR region | - | Moderate (37) | High (44) | Low (27) | Can be used as a feature to distinguish K deficiency from N and Mg deficiencies. But N and Mg deficiencies may not be uniquely identified. |
| Brightness at 550 nm | Lowest (~4614) | Highest (~16000) | Lower (~9707) | Lower (~12126) | Can be used as a feature to distinguish control and N nutrient deficiency. But K and Mg deficiencies may not be uniquely identified. |
| Brightness at 800 nm | Highest (~16210) | Lower (13820) | Lowest (12700) | Lower (14616) | Can be a significant feature to distinguish control from nutrient deficient leaves. But an individual nutrient deficiency may not be uniquely identified. |

| Ratio (Control /Nutrient Deficiency) | - | Highest at ~ 617 nm (3.7015) | Low at ~617 nm (2.16636) | Moderate at ~617 nm (2.743) | Can be used as a feature to identify control and an individual nutrients' deficiency. |
|---|---|---|---|---|---|
| SD in UV region | Lowest (~1500) | High (~2300) | Lower (~1800) | Lower (~1700) | Can be used as a feature to identify control leaves. May also be used for N deficient leaves' detection. |
| SD in VIS region | Lowest (~1000) | High (~4100) | Lower (~2000) | Lower (~2600) | Can be used as a feature to identify control and an individual nutrients' deficiency. |
| SD in NIR region | Highest (~2200) | Lower (~1300) | Low (~1200) | High (~1600) | Can be used as a feature to detect control leaves. |
| 1st derivative ratio between 680 nm and 750 nm | ~ 30% Positive direction | ~ 62% Negative direction | ~ 37% Negative direction | ~47% Negative direction | Can be used as a feature to identify control and an individual nutrients' deficiency. |

**5. Conclusions and Future Work**

This paper presented an experimental study of age detection and distinguished between healthy and unhealthy grapevine leaves with the aid of HS imaging in UV, VIS and NIR wavelength ranges (~380 nm to 1000 nm). It suggests that this technology could be efficiently employed to obtain reliable results in this domain.

*5.1. Age Detection*

For age detection, leaves aged very young to old have been studied. From the analysis, it was found that the magnitude of the mean brightness of leaves has a strong correlation with ageing. This variation was found to be more prominent at ~ 554.3 nm. Mean 1st derivative brightness study demonstrated that mean 1st derivatives brightness peaks in VIS wavelengths shifted to the longer wavelengths as the leaves age. Additionally, mean 1st derivatives brightness curves in the range between ~ 675 nm and ~ 775 nm revealed that the rate of change of mean brightness decreases with gradual ageing of leaves. Variation index, a new metric introduced in this manuscript, also indicated to be correlated with age of leaves.

*5.2. Healthy/Unhealthy Leaves Detection*

Leaf samples were divided into two groups to distinguish between healthy and unhealthy leaves. The 1st group contained leaves that were visibly unhealthy, and features of these leaves were compared with those of visibly health leaves (i.e. benchmark leaf). In this case, both the whole area and selective areas containing defects on the leaves were investigated. It was found that the mean 1st derivatives brightness in the NIR region disclosed distinguishable curves for both whole area and selective areas compared to a healthy leaf. But the variation of curve shapes was more significant for whole area leaves compared to selective area leaves. Furthermore, the ratio analysis study established that comparably more distinguishable curves could be obtained from the whole leaf study. Experimental results suggest that the variation index could be employed to detect defective leaves effectively.

*5.3. Nutrient Deficiency Detection in Leaves*

The 2[nd] group of grapevine leaves was incorporated in our analysis to extend our understanding of specific nutrient deficiencies in leaves. These leaves had N, K or Mg deficiencies individually. The features of these leaves were studied relative to those of control leaves and were also compared between each other.

It was found that the brightness feature in the VIS range can be used to distinguish between control and nutrient deficient leaves as the brightness of leaves with nutrient deficiencies is higher than that of control leaves. This characteristic can also be employed to identify control and N deficient leaves due to the highest brightness peak of N deficient leaves compared to control, K and Mg deficient leaves at ~ 550 nm. Brightness peaks of nutrient-deficient leaves appeared at the shorter wavelength (i.e. redshift) compared to the control leaves in the wavelength range between ~ 680 nm and 1000 nm.

Furthermore, results suggest that NDVI values could significantly differentiate control and nutrient deficient leaves though N, K and Mg deficiencies could not be identified uniquely.

Various features of nutrient-deficient leaves between ~ 650 nm to 800 nm of 1[st] derivative peaks compared to the control leaves could be useful to distinguish between control and nutrient deficient leaves. The attributes such as dominant peak at ~700 nm, narrowing of 1[st] derivative brightness peaks and 1[st] derivative peaks shifted to the shorter wavelength (redshift) are noteworthy.

1[st] derivative mean brightness ratio plots between 680 nm and 750 nm could be employed to distinguish control, and nutrient-deficient leaves as this ratio are positive for control leaves and negative for nutrient deficient leaves. Furthermore, the magnitude of this ratio of different nutrient-deficient leaves could be utilised to identify leaves with N, K and Mg deficiencies.

The study of SD in UV, VIS and NIR wavelength regions states that this feature could be used to distinguish control and individual nutrients' deficient leaves as different SD values were obtained for N, K and Mg deficiencies.

Additionally, N and K deficiencies could be identified from the variation index analysis in UV and NIR wavelengths, respectively.

*4.4. Future Work*

In future we envisage analysing all or most of the Micro and Macronutrients that grapevines require including but not limited to:
- Micro-nutrients: copper (Cu), iron (Fe), manganese (Mn), boron (B), chlorine (Cl).
- Macro-nutrients: calcium (Ca), oxygen (O), sulfur (S), phosphorus (P), carbon (C).

There might also be some additional theoretical analysis presented in the future. A mathematical model might be developed for all age groups. The result of modelling can be used in classification as well as future prediction of more significant deficiency levels of various leaf age groups. It might also be interesting to experiment whether similar findings can be obtained for leaves of other species.